\documentclass[10pt]{article}
\usepackage{amssymb}
\begin{document}         
\title{Quantum Physics and Classical Physics - in the Light of Quantum
Logic}
\author{Peter Mittelstaedt, University of Cologne, Germany}

\maketitle

\begin{abstract}
  In contrast to the Copenhagen interpretation we consider  quantum mechanics as
  universally valid and query whether  classical physics is really intuitive and plausible. - We
  discuss these problems within the quantum logic approach to quantum mechanics where the
  classical ontology is relaxed by reducing metaphysical hypotheses. On the basis of this weak 
  ontology   a formal logic of quantum physics can be established which is given by an
  orthomodular lattice. By means of  the Sol\`er condition and Piron's result one obtains the
  classical Hilbert spaces. - However, this approach is not fully convincing. There is no
  plausible justification of Sol\`er's law and  the quantum ontology is partly too weak and partly
  too strong. We propose to replace this ontology  by an ontology of unsharp properties and 
  conclude that quantum mechanics is more intuitive than classical mechanics and that classical
  mechanics is not the macroscopic limit of quantum mechanics.
\end{abstract}

\section{\mbox{The dualism of  Copenhagen interpretation}}

  Even today, 75 years after the discovery of quantum mechanics many quantum
  physicists are convinced that the Copenhagen interpretation  is still the
  right way for understanding quantum physics. According to this interpretation we
  have to distinguish two distinct worlds, the quantum world of microscopic entities and
  the classical world of our everyday experience which is subject to classical physics. In
  the quantum world we are confronted with many strange features, complementarity,
  nonindividuality, nonlocality, and the loss of determinism. However, the apparatuses
  which measure and register the properties of the quantum system as well as the human
  observer, who reads the observed data are parts of the classical world that is free from
  the quantum physical absurdities mentioned. For describing and interpreting quantum
  physics we can use common language  and classical logic. 
  
  We will query this doctrine here for several reasons. Firstly, during the last decades it
  became obvious that quantum mechanics is not restricted to the microscopic world of
  nuclei, atoms, and molecules but can be applied also to macroscopic systems. The
  discovery of macroscopic quantum effects like superconductivity, superfluidity,
  macroscopic tunnelling, etc. are strong indications that quantum physics holds also in
  the macroscopic world. Moreover, the successful attempts to a quantum cosmology 
  show that quantum mechanics can even be applied on the cosmological level, to the
  problem of the creation of the universe and to the universe as a quantum mechanical
  object. Hence it seems, that there are no serious doubts today that quantum mechanics
  is universally valid and can be applied to all objects from elementary particles to the
  entire universe. 
  
  The second reason is presumably even more important. In the Copenhagen
  interpretation quantum mechanics is considered to be less intuitive than classical
  mechanics and sometimes paradoxical, whereas classical mechanics is assumed to
  correspond to plausible reasoning and to intuitive results. This is, however, not
  entirely correct. What we call  ``intuitive'' and in accordance with  ``plausible
reasoning''
  corresponds to our everyday experience, to our pre-scientific experience in the
  macroscopic world. However, classical physics and in particular classical mechanics
  is not exactly the theory of this pre-scientific experience. Classical mechanics is
  loaded with many hypotheses which can be traced back to the metaphysics of the
  17$^{\mbox{\tiny th}}$ and 18$^{\mbox{\tiny th}}$ century.  These
  metaphysical hypotheses are without any empirical counterpart, they exceed clearly
  our everyday experience. As examples we mention here the existence of an absolute
  time, the complete determination of objects, the strict causality law, and the law of
  conservation of substance. It is obvious that the consequences of these  hypotheses 
  are not {\em per se} intuitive in the above mentioned sense. 
  
  In quantum mechanics we are confronted with a quite different situation. Quantum
  mechanics may be understood as a theory of  the physical reality which is free from
  some of the metaphysical hypotheses mentioned, i.e. quantum mechanics dispenses
  with some metaphysical exaggerations of the classical theory. It is important to note
  that quantum mechanics can be obtained from classical mechanics merely by reducing 
  the ontological premises without incorporating  new empirical components. This will
  be demonstrated in detail within the framework of the quantum logic approach to
  quantum mechanics. Consequently, in quantum mechanics just those parts of classical
  mechanics are missing which are not intuitive  and which do not correspond to
  plausible reasoning. This means that quantum mechanics is more intuitive than
  classical mechanics -- a result which is paradoxical at first glance. It is obvious that
  this result together with the universal validity of quantum mechanics strongly
  invalidates the dualistic approach of the Copenhagen interpretation.  

  \section{Aiming a new quantum ontology}

  The ontology of a certain domain of physics contains the most general features of the
  external reality which is treated in the physical domain in question. In particular the
  ontology should contain the material preconditions for a  pragmatics which allows for
  the constitution of a scientific language  and thus for the formulation of physical
  experience. The ontology which is underlying classical mechanics will be called
  classical ontology and denoted by $O(C)$. We will briefly characterise this classical
  ontology.
  
  According to $O(C)$ there are individual and distinguishable objects $S_i$ and these
  objects possess elementary properties $P_\lambda$ in the following sense. An elementary
  property $P_\lambda$ refers to a classical object system such that either
  $P_\lambda$ or the counter property $\bar P_\lambda$ pertains to the system. 
  An elementary property $P_\lambda$ can always be tested
  by measurement with the result that either $P_\lambda$ or the counter property
  $\bar P_\lambda$ pertains to
  the object. Furthermore, objects are subject to the law of ``complete
  determination''
  according to which  ``if all possible predicates are taken together with their
  contradictory opposites then one of each pair of contradictory opposites must belong
  to it'' [Kant, 1920]. Hence an object $S$ possesses each elementary property
  $P$ either
  positive $(P)$ or negative $(\bar P )$. It follows from these strong requirements that objects
  can be individualised by elementary properties if impenetrability is assumed as an
  additional condition. For objects of the external objective reality the causality law and the
  law of  conservation of substance hold without any restriction. Since there exist an absolute
  and universal time which refers to all objects of the external reality, the temporal
  development of these objects and their time dependent properties are strictly
  determined by a causal law of nature which fulfils also the conservation of substance. 
  
  There are important objections against this classical ontology. Since the metaphysical
  and theological reasons of  Newton are no longer relevant for a justification of the
  ontology we have to search for alternative reasons. Are the ontological assumptions
  intuitive and plausible in the sense mentioned above? This is obviously not the case.
  The strict postulates of the classical ontology are almost in accordance with our
  everyday experience, but the rigorousness of the assumptions mentioned exceeds
  obviously the more qualitative and less rigorous prescientific everyday experience.
  The strict causality law, the unrestricted conservation of substance and the existence
  of  one universal time are beyond  our daily experience. These and other hypotheses
  of the classical ontology must not be considered  as intuitive and plausible. 
  
  The second argument refers to the experimental evidence of the mentioned
  hypothesis. There is no experimental indication that objects can always be
  individualised and reidentified at later times, simply since experiments which would
  confirm this assumption have never been performed in classical physics. In addition,
  the principle of complete determination mentioned above has never been tested with
  an accuracy which would allow to call the result a principle. Consequently, there is no
  justification for a strict causality law such that the present state of an object allows for
  predictions about all elementary properties. Hence we find that there is no empirical
  justification for the classical ontology  $O(C)$. Instead, the classical ontology is based
  on hypotheses whose origin can be traced back to the metaphysics of the
  17$^{\mbox{\tiny th}}$ and 18$^{\mbox{\tiny th}}$
  century. 
  
  The classical ontology is neither intuitive and plausible nor is it justified by
  experimental evidence. Moreover, -- what is more important -- the classical ontology
  $O(C)$ is not in accordance with quantum physics. A quantum mechanical object
  system does not possess all possible elementary properties $P_\lambda$ either positive
  $(P_\lambda )$ or negative $(\bar P_\lambda )$. 
  It is not carrier of all possible properties. Instead, only a subset of all
  properties pertains to the system and can simultaneously be determined. These
  properties are often called   ``objective'' properties and they pertain to the object like in
  classical ontology.  From these restrictions it follows that in quantum mechanics no
  strict causality law can be  established and that object systems cannot be
  individualised and reidentified by means of their objective properties.  
   
  We will not use here these empirical results for a reconstruction of an ontology for
  quantum phenomena. However, we learn from these considerations, that the classical
  ontology is not only based on classical metaphysics and partially hypothetical, but
  that classical ontology contains too much structure and too strong requirements
  compared with quantum physics. This observation offers the interesting possibility to
  formulate the ontology $O(Q)$ of quantum physics by relaxing and weakening some
  hypothetical requirements of the classical ontology $O(C)$. It is important to note that
  no new requirements must be added to the assumptions of the classical ontology.
  Quantum ontology can thus be formulated  as a reduced version of the classical
  ontology $O(C)$:
  \begin{itemize}
  \item[$O(Q)$-1:] If an elementary property $P$ pertains to a system as an objective property,
  then a test of this property by measurement will lead with certainty to the result
  $P$. In
  addition, any arbitrary elementary property $P$ can be tested at a given object with the
  result that either $P$ or the counter property $\bar P$  pertains to the object system. (These
  requirements are in complete accordance with $O(C)$). 

\item[$O(Q)$-2:] Quantum objects are not completely determined. They possess only a few
  elementary properties either positive or negative. Properties which pertain
  simultaneously to an object are called  ``objective'' and ``mutually
commensurable''.

\item[$O(Q)$-3:] For quantum objects there is no strict causality law, simply since the present
  state of an object system is never completely determined. 
  
\item[$O(Q)$-4:] The lack of complete determination and of strict causality implies that
  quantum objects cannot be individualised and reidentified at later times.  

\end{itemize}
  
  The mutual relations between classical and quantum ontology are the key for the very
  intertheoretical relations between classical and quantum physics. This will become
  obvious  in the quantum logic approach to quantum mechanics when quantum
  ontology is used as starting point for establishing a formal language and logic of
  quantum physics. This way of reasoning will be made explicit in the following
  section.

\section{The quantum logic approach to quantum mechanics}
  
  \subsection{Language, semantics, and pragmatics}
       On the basis of the quantum ontology $O(Q)$ described above we will establish
  a formal language of quantum physics. Let $S$ be a proper quantum system and
  $A, B, \dots$ elementary propositions which attribute predicates (properties)
  $P(A), P(B), \dots$ to system $S$ at times $t_1, t_2, \dots$
  Hence, we write for  elementary propositions $A(S, t_1), B(S,t_2), \dots$ 
  According to  $O(Q)$-1 we assume that for every elementary proposition $A$ there
  exists a finite testing procedure which shows whether $P(A)$ pertains to
  $S$ or not.  If 
  $P(A)$ pertains to $S$ at time $t_1$ , then the proposition $A(S, t_1)$ is called to be
 ``true'',
  otherwise $A(S, t_1)$ is said to be  ``false''. The assumption, that for every elementary
  proposition there is a testing procedure which decides between  ``true'' and
  ``false''
  means, that these propositions are  ``value definite''. Hence, an elementary proposition
  can either be proved (with result $A$) or disproved (with  result $\bar A$),  where
  $\bar A$ is the
  counter proposition of $A$. 
       Furthermore, we assume that after a successful proof of  $A$ new proof attempt
  leads with certainty to the same result, provided the time interval between the two
  proof attempts is sufficiently small. This requirement is again in accordance with
  $O(Q)$-1. Since after the first test the property $P(A)$ pertains objectively to the system
  and can thus be tested with the certain result $P(A)$. This assumption means that there
  are repeatable measurement processes, which can be applied to the testing procedures.
  However, -- and this is an important restriction of $O(C)$ -- if after a successful proof of
  $A$, say, another proposition $B$ is proved, then a new proof attempt of
  $A$ will in general
  not lead to the previous result. Hence, we will not assume that two propositions
  $A$ and
  $B$ are in general simultaneously decidable. If accidentally two propositions
  $A$ and $B$
  are always jointly decidable, we will call $A$ and $B$ ``commensurable''. In this case,
  after the proof attempt of $B$ the result of the previous $A$-test is still available.
  However, in the general case the result of a previous test is only restrictedly available. 
       
  On the basis of the set  ${\cal S}_Q{}^{\rm e}$  of elementary propositions we introduce the
  logical connectives by the possibilities to attack or to defence them, i.e. by the
  possibilities to prove or to disprove the connective. Here, we consider the sequential
  conjunction $A\sqcap B$ ($A$ and then $B$) which refers to two subsequent instants of time
  $t_1$ and
  $t_2$  with  $t_1 < t_2$  and the logical connectives $\neg A$ ( not A),
  $A\wedge B$ (A and B), $A\vee B$ ($A$ or
  $B$), and $A\rightarrow B$ (if $A$ then $B$) -- which refer to one simultaneous instant of time.  The
  definitions of the sequential and logical connectives by attack-and defence schemes
  can be illustrated most conveniently by chronologically ordered proof trees.
  Correspondingly, in the proof tree of the sequential conjunction $A\sqcap
  B$, the first
  branching point corresponds to a $A$-test at $t_1$, the second one to a
$B$-test at $t_2$. 
       Note, that for the truth of $A \sqcap B$ the commensurability of $A$ and
$B$ does not
  matter. However, for the proof trees of the logical connectives, which refer to one
  simultaneous instant of time, the commensurabilities of the elementary propositions
  play an important role. The concepts of truth and falsity of a compound proposition
  which is composed by the connectives can then be defined by success and failure in a
  proof tree, respectively. [Mittelstaedt, 1978; Stachow, 1980; Mittelstaedt,  1987].

       Furthermore, we will define here binary relations between propositions. First,
  the proof equivalence $A\equiv B$ means that $A$ can be replaced in any proof tree of a
  compound proposition by $B$ without thereby changing  the result of the proof tree.
  Second, the value equivalence $A = B$ means that $A$ is true (in the sense of a proof tree)
  if and only if $B$ is true. Third,  the relation of implication $A\leq B$  can be defined by 
  $A\equiv A\wedge B$. Hence, the two  implications $A\leq B$ and $B\leq A$ imply the proof equivalence
  $A\equiv B$. Finally, we mention that $A\rightarrow B$ is true if and only if
  $A\leq B$ holds. 
       The full quantum language ${\cal S}_Q$ can then inductively be defined by the set
  ${\cal S}_Q{}^{\rm e}$  of
  elementary propositions  and the connectives mentioned. Together with the always
  true elementary proposition ${\rm V}$, the always false elementary proposition
  $\Lambda$, and the three relations one obtains  the language ${\cal S}_Q$.

\subsection{Quantum logic}
  
       The semantics described here is a combination of a realistic semantics (for
  elementary propositions) and a proof semantics (for connectives). Hence, the truth of
  a compound proposition depends on the connectives contained in it as well as on the
  elementary propositions and their truth values. However, there are finitely connected
  propositions which are true in the sense of the semantics mentioned, irrespective of
  the truth values of the elementary propositions contained in it. These propositions are
  called  {\em formally true}. 
  -- The precondition that measurements are repeatable implies that 
  $A\rightarrow A$, the {\em law of identity}, 
  is formally true. The value definiteness of elementary
  propositions implies that also finitely connected propositions are value definite and
  thus  $A\vee\neg A$, the {\em tertium non datur law}, is formally true. 
  In a similar way, it follows
  that $\neg(A\wedge\neg A)$, the {\em law of contradiction}, and
  $(A\wedge(A\rightarrow B))\rightarrow B$, the modus ponens law,
  are formally true. -- Formally true propositions can also be expressed by
  ``formally true
  implications''. E.g. the modus ponens law reads $A\wedge(A\rightarrow B)
  \leq B$. In addition, the relations  
  $A\leq {\rm V}$  and  $\Lambda \leq A$ hold for all propositions $A\in
  {\cal S}_Q$. The formal truth of a proposition $A$
  can then be expressed by ${\rm V} \leq A$.  E.g. the tertium non datur law reads
  ${\rm V} \leq A\vee \neg A$ .

       There are two kinds of propositions $A\in{\cal S}_Q$. If a compound proposition contains
  in addition to elementary propositions only the logical
  connectives $\wedge,\vee,\neg$ and $\rightarrow$ , then it is called a  ``logical
  proposition''. In the more
  general case, when the proposition contains also {\em sequential} connectives, 
  in particular
  the sequential conjunction $\sqcap$ , then it is called a ``sequential
  proposition''. In addition to
  the formally true logical propositions mentioned above, there are also formally true
  sequential propositions. If $A$ and $B$ are logical propositions then
  $A\wedge B \leq A\sqcap B$ is a
  formally true implication. The totality of  formally true implications can be
  summarised in a calculus which contains  ``beginnings''   $\Rightarrow A
  \leq B$  and rules $A \leq B \Rightarrow  C \leq D$. 
  Here, we distinguish the calculus ${\bf L}_Q$ of formally true {\em
  logical}  propositions and
  the calculus ${\bf S}_Q$ of formally true sequential propositions. [Mittelstaedt, 1978; Stachow,
  1980].

       For an algebraic characterisation of the calculi ${\bf L}_Q$ and
  ${\bf S}_Q$ we consider the
  corresponding Lindenbaum-Tarski algebras.  The
  Lindenbaum-Tarski algebra of the calculus ${\bf L}_Q$ is given by a complete, orthomodular
  lattice $L_Q$. Subsets of mutually commensurable propositions constitute a Boolean
  sublattice  $L_B \subseteq  L_Q$ of the  lattice $L_Q$ [Mittelstaedt,
  1987]. Moreover, if the entire
  quantum language ${\cal S}_Q$ refers to one individual quantum system, then the lattice
  $L_Q$ is
  atomic and fulfils the covering law
  [Stachow, 1984].  In this case we denote the lattice by  $L_Q^*$.  The Hilbert lattice
  $L_H$ of
  projection operators in Hilbert space can be
  obtained from the lattice  $L_Q^*$   by adding the Sol\'er law, the meaning of which is,
  however, still open [Sol\'er, 1995]. Correspondingly, the Lindenbaum-Tarski algebra of
  the calculus ${\bf S}_Q$ of sequential quantum logic is given by a
  Baer$^*$ semigroup. [Stachow, 1980;  Foulis,1960]. It is
  well known that by means of  a result by {\em Piron} [Piron, 1976] from the lattice
  $L_H$ the
  three classical Hilbert spaces can be obtained and that for the complex
  numbers~$\mathbb{C}$
  quantum mechanics in Hilbert space is achieved.

\subsection{Is quantum mechanics a priori valid?}
  
  The described approach to quantum mechanics which starts from the
  relaxed quantum ontology $O(Q)$ and leads finally to the quantum mechanical Hilbert
  space, is sometimes considered as an a priori justification of quantum theory
  [Mittelstaedt, 1978]. The term  ``a priori'' seems to be legitimated here, since the
  starting point of this approach are the most general preconditions of a scientific
  language of physics, i.e. the assumptions of the weak ontology $O(Q)$. However, this
  way of reasoning is not fully convincing. Firstly, up to now there is no plausible and
  intuitiv justification of Sol\'er's law, which appears in the present approach as an
  additional {\em ad hoc} assumption. Hence, one could ask whether the quantum ontology
  $O(Q)$ is really the right starting point.
  $O(Q)$ is too weak, since the main restriction of quantum ontology with respect to
  classical ontology, the complementarity requirement, is a very strong postulate. Two
  properties which are not commensurable for accidental reasons are
  complementary in the sense that they cannot be tested simultaneously.
  Complementarity in this strong form must be required in quantum mechanics for
  sharp observables which are given by PV-measures.

  However, even in quantum mechanics the strong complementarity requirement can be
  relaxed by the uncertainty principle making use of unsharp observables in the sense of
  POV-measures. POV-measures are the most general observables which allow for a
  probability interpretation 
  of quantum mechanics [Busch et al. 1995]. Two unsharp properties of a quantum
  system can be attributed jointly to the object, if the conveniently defined degrees of
  unsharpness  of the two properties fulfil the Heisenberg uncertainty inequality
  [Busch, 1985]. Obviously, a quantum ontology $O(Q^{\mbox{\scriptsize\rm u}})$ which
   replaces the complementarity
  requirement by the uncertainty principle, is somewhat stronger than the original
  ontology $O(Q)$. 
  
  The ontology $O(Q)$ is not only too weak but -- with respect to another feature
  -- also too
  strong. In accordance with the classical ontology $O(C)$ we assumed in
  $O(Q)$-1 that any
  elementary property $P$ can be tested experimentally with the result that either
  $P$ or the
  counter property $\bar P$ pertains to the system. This ontological precondition implies that
  elementary propositions of the quantum language are value definite and that the
  tertium non datur holds in quantum logic ${\bf L}_Q$ . However, the
  ontological precondition that any elementary property can be tested by experiment
  (with the result $P$ or $\bar P$) exceeds the possibilities of Hilbert space quantum mechanics.
  Within the framework of the quantum theory of  unitary premeasurements it follows
  that pointer objectification cannot be achieved for closed systems [Mittelstaedt
  1998].
  Hence, value definiteness of elementary propositions is incompatible with quantum
  mechanics in Hilbert space and must be relaxed in some sense. In this situation it
  suggests itself to begin with elementary propositions that are not value definite
  and correspond to unsharp properties 
  given by POV-measures\footnote{It must be mentioned that up to now it is
  not yet quite clear whether the problem of pointer objectification can
  completely be solved by POV-measures [Busch 1998].} 
  Hence it seems that also the second objection against the
  quantum ontology $O(Q)$ can be taken account of by the quantum ontology
  $O(Q^{\mbox{\scriptsize\rm u}})$
  based on unsharp properties.

  Hence, on the basis of the slightly modified quantum ontology
  $O(Q^{\mbox{\small\scriptsize u}})$  a fresh start  by
  means of unsharp properties seems to be a quite promising attempt. In a first step of
  this approach a formal language and logic of not necessarily value definite quantum
  mechanical propositions must be developed. In a second step from the algebraic
  structure of the logic the algebra of effects and the Hilbert space must be
  reconstructed. Hence one could either start from a language of unsharp propositions or
  from a modified algebraic structure of  quantum logic. In recent years many
  interesting logical systems for unsharp propositions were proposed. 
  [Dalla Chiara 1993, 1994, 1995, Foulis 1997, Giuntini 1989, 1990, Mittelstaedt 1978]

However, it is still an open question whether in this way a consistent operational 
  approach to quantum mechanics can be obtained. Up to now the logical systems
  mentioned were not yet reconstructed in an operational way starting from a formal
  language of unsharp propositions. Furthermore, the Lindenbaum-Tarski algebra of a
  logical calculus of unsharp propositions is not {\em per se} equivalent to the algebra of
  effects in Hilbert space. We do not know which kind of law must be added to the
  algebra of unsharp propositions in order to obtain the effect algebra mentioned (it
  could be as complicated as the Sol\'er law). Even the last step of an operational
  foundation of quantum physics, the way from the effect algebra to the Hilbert space
  requires more detailed investigations.

  \subsection{The ontological priority of quantum mechanics}
  
  Although the task of  reconstructing quantum mechanics on the basis of a logic of
  unsharp propositions is not yet finally performed, we can draw already some
  interesting conclusions which refer to the interpretation of quantum mechanics. The
  basis of this approach, the (uncertainty) ontology $O(Q^{\mbox{\scriptsize \rm u}})$ is somewhat richer than the
  (complementarity) ontology $O(Q)$ but weaker than the classical ontology
  $O(C)$ of
  complete determination. This classical ontology is not only based on experience but
  also on several metaphysical hypotheses -- which are weakened or cancelled in
  $O(Q^{\mbox{\scriptsize \rm u}})$.

  Since these metaphysical hypotheses (complete determination, individuality, and full
  determinism) clearly exceed our everyday experience, and since we call phenomena
  intuitive and understandable if they are in accordance with this everyday experience,
  classical mechanics is not thoroughly intuitive. However, since the hypotheses
  contained in the classical ontology $O(C)$ are strongly reduced in the (uncertainty)
  ontology $O(Q^{\mbox{\scriptsize \rm u}})$ as well as in the (complementarity) ontology
  $O(Q)$, we expect that
  the implications of the quantum ontology $O(Q^{\mbox{\scriptsize \rm u}})$ are more intuitive and more plausible
  than the implications from the classical ontology.

  In particular the quantum logic approach can further illustrate this result. The logical
  systems  
  which follow  from the ontologies $O(Q)$ and $O(Q^{\mbox{\scriptsize \rm u}})$ are based on weaker and less
  hypothetical pragmatic preconditions than the Boolean lattice $L_B$ of classical logic.
  Hence the resulting quantum mechanics is more intuitive and more plausible  than
  classical mechanics. In addition, since quantum mechanics is based on weaker
  premises than classical mechanics, it is  nearer to the ``truth'' than classical mechanics.

  On the basis of these results we can formulate the r\^{o}le of classical mechanics.
  {\em Firstly},
  classical mechanics is loaded with metaphysical hypotheses which clearly exceed our
  everyday experience. Since quantum mechanics is based on strongly relaxed
  hypotheses of this kind, classical mechanics is less intuitive and less plausible than
  quantum mechanics. Hence classical mechanics, its language and its logic cannot be
  the basis of an adequate interpretation of quantum mechanics. {\em
  Secondly}, classical
  mechanics is not the limit of quantum mechanics for macroscopic phenomena. Since
  quantum mechanics of closed systems does not explain the objectification, i.e. the
  classical behaviour of pointer values, classical mechanics cannot be the macroscopic
  limit of quantum mechanics. However, this argument which is still subject of
  controversial debates, is not the main reason. The essential argument which shows
  that classical mechanics is not the limiting case of quantum mechanics  is based on
  the observation that classical mechanics is loaded with metaphysical hypotheses
  without any empirical counterpart. Since some of these hypotheses are explicitly
  eliminated in quantum theory, it is obvious that there is no approximation procedure
  which leads from quantum mechanics to classical mechanics. Classical mechanics
  describes a fictitious world which does not exist in reality.

\end{document}